\documentclass[usenatbib,usegraphicx]{mn2e}

\title[Rotation periods for TW Hydrae association stars]
{Rotation periods for stars of the TW Hydrae association:
the evidence for two spatially- and rotationally-distinct 
pre-main sequence populations}

\author[W. A. Lawson \& L. A. Crause]
{Warrick A. Lawson$^{1}$\thanks{E-mail: w.lawson@adfa.edu.au 
(WAL); lisa@saao.ac.za (LAC)} and Lisa A. Crause$^{2, 3\star}$\\
$^{1}$School of Physical, Environmental and Mathematical Sciences,
The University of New South Wales,\\
$^{~}$Australian Defence Force Academy, Canberra ACT 2600,
Australia\\
$^{2}$Department of Astronomy, University of Cape Town, Private Bag,
Rondebosch 7700, South Africa\\
$^{3}$South African Astronomical Observatory, P. O. Box 9, 
      Observatory 7935, South Africa}

\begin{document}

\date{Accepted January 2005}

\pagerange{\pageref{firstpage}--\pageref{lastpage}}
\pubyear{2005}

\maketitle

\label{firstpage}

\begin{abstract}
We have conducted a photometric study of late-type members of the 
TW Hydrae association (TWA) and measured the rotation periods for 
16 stars in 12 systems.  For TWA stars listed by Webb et al. and
Sterzik et al. (TWA $1-13$; led by TW Hya = TWA 1) we find a median 
period of 4.7 d.  However, for stars that we measured in the TWA 
$14-19$ group identified by Zuckerman et al., we find a median 
period of only 0.7 d.  The period distributions of the two groups 
cannot be reconciled at the 3-$\sigma$ significance level.  Using 
photometric arguments supported by the {\it Hipparcos\,} distance 
to HD 102458 (= TWA 19A), we find that TWA $14-19$ reside at an
average distance of $d \approx 90$ pc, spatially at the near 
boundary of the Lower Centaurus-Crux (LCC) subgroup of the 
Ophiuchus-Scorpius-Centaurus OB-star association (OSCA).  Proper
motions for HD 102458, TWA 14, 18 and 19B link these stars to the
LCC subgroup.  From Hertzsprung-Russell (H-R) diagram placement, 
we derive an age of $\approx 17$ Myr for the HD 102458 system that 
may be the representative age for the TWA $14-19$ group.  Merging 
various lines of evidence, we conclude that these stars form a
spatially- and rotationally-distinct population of older pre-main
sequence (PMS) stars, rather than being an extension of the TWA 
beyond those stars associated with TW Hya that have an age of 
$\sim 10$ Myr and reside at $d \approx 55$ pc.  Instead, TWA 
$14-19$ likely represent the population of low-mass stars still
physically associated with the LCC subgroup. 
\end{abstract}

\begin{keywords}
stars: pre-main sequence ---
stars: rotation ---
stars: activity ---
open clusters and associations: individual: TW Hydrae
\end{keywords}

\section{Introduction}

The TWA remains one of the most extensively studied PMS populations,
owing to the novelty of the association's namesake; TW Hydrae is a
classical T Tauri (CTT) star remote from the nearest sites of star
formation \citep{Herbig78}.  The proximity of the association to 
Earth and the brightness of its component stars and brown dwarfs
enables high sensitivity spatial, photometric and spectroscopic 
study at all wavelength regimes, e.g. the $V = 10.8$ TW Hya has 
an {\it Hipparcos\,} distance of $d = 56$ pc.  The TWA is also at
the astrophysically important age of $t \sim 10$ Myr, comparable 
to the timescales for the end of disc accretion 
\citep{Muzerolle00, Lawson04}, and the dissipation of circumstellar 
discs and the incorporation of disc material into planetisimals.  
Disc dissipation is also believed to be related to the angular 
momentum evolution of PMS stars as they `unlock' from their discs 
at the end of the accretion phase and spin up as they evolve down 
their Hayashi tracks; for a recent study, see Rebull, Wolff \& 
Strom (2004).

The history of the discovery of other T Tauri stars in the region
of TW Hya is documented by \citet{Webb99} who defined the TWA as
including 11 star systems containing at least 19 stars and one 
brown dwarf spanning spectral types of A0 to M8.5. From X-ray and
kinematic study, surveys have since added other stars purported to
be TWA members (Sterzik et al. 1999; Zuckerman et al. 2001a; 
Reid 2003; Song, Zuckerman \& Bessell 2003), bringing the TWA 
membership to 25 systems, despite speculation as to whether all 
of these T Tauri systems truly form a physical association; e.g.
\citet{Jensen98, Song03}.  

By back-tracking space motions for some of the brighter TWA stars
with accurate distances and kinematics, the TWA appears to be an 
outlying population of the LCC (or perhaps the Upper Centaurus-Lupus) 
subgroup of the OSCA, possibly formed within small molecular clouds
that were collected by expanding supernova bubbles $\sim 10$ Myr 
ago (Mamajek, Lawson \& Feigelson 2000; Mamajek, Meyer \& Liebert
2002).  The TWA is one of several nearby PMS groups of similar age 
($5-12$ Myr), but $\sim 10$ Myr younger than the OSCA OB-star 
subgroups \citep{Mamajek02}, that are thought of as dispersed groups
of OSCA origin.  Other outlying populations include the $\eta$ Cha
(Mamajek, Lawson \& Feigelson 1999) and $\epsilon$ Cha clusters 
(Feigelson, Lawson \& Garmire 2003) and the $\beta$ Pic moving 
group \citep{Zuckerman01b}.

Whatever the concern over the homogeneity or origin of the TWA,
there is intense interest in this group of 
nearby T Tauri stars.  However the available suite of fundamental 
information on TWA members is still incomplete.  Within a study 
based on published rotational information for the stellar populations 
of young clusters by \citet{Rebull04}, a notable omission was the 
lack of rotation periods for members of the TWA, except for TW Hya 
itself.  In this paper we in-part remedy this situation with the 
measurement of rotation periods for 16 TWA stars in 12 TWA systems, 
the outcome of a multi-epoch multi-colour photometric survey of 
many of the TWA systems.

\section{Observations and Data Reduction}

An initial trial of multi-epoch $V$-band observations for many 
of the TWA stars listed by \citet{Webb99} and \citet{Sterzik99}
was made using the 1-m telescope and 1k $\times$ 1k SITe 
charge-coupled device (CCD) at the Sutherland fieldstation of 
the South African Astronomical Observatory (SAAO) over 3 weeks
of observing time during 2000 February and March.  The acquisition 
of these observations was secondary to a programme to measure 
rotation periods for stars of the $\eta$ Chamaeleontis cluster 
\citep{Lawson01} and for that reason only $\approx 15$ measurements 
were obtained for each TWA star, an insufficient number for a 
definitive evaluation of the rotation periods in many cases.  To
address this outcome, a 2-week observing run from 2001 May $1-14$
using the same instrumentation was dedicated to the TWA stars, 
including several additional members identified by 
\citet{Zuckerman01a}.  For most of the run the weather was ideal
for differential photometry, with observations being obtained on 
11 nights of which the final 9 were consecutive.

All of the TWA stars observed were measured at the $V$- and Cousins
$I$-bands, with 3 stars (TWA 1 = TW Hya, TWA 6 and TWA 12) also 
observed at $B$-band based upon the preliminary 2000 results that 
suggested either an irregular light curve (TW Hya) or high-amplitude 
light curves (TWA 6 and 12).  For most of the stars, $2-4$ 
observations were made each night to reduce the 1-day aliasing errors
in the light curve analysis that might be introduced from the 
single-site observations.  Most stars were observed on $\approx 30$
occasions, while TW Hya was observed $\approx 40$ times, and TWA 6 
was observed $\approx 70$ times.  Not all TWA stars were observed; 
several were too bright for the 1-m telescope and/or resided in 
sparse fields without suitable comparison stars within the small 
(5.3 arcmin) field-of-view of the CCD, or they consisted of binary 
systems with arcsec-level separations that were unresolved in the 
$1-2$ arcsec seeing conditions typical for the observing run.  Based 
on the membership lists of \citet{Webb99}, \citet{Sterzik99} and
\citet{Zuckerman01a}, we did not observe TWA 4AB, 11AB and 19AB.  
Of the 24 stars we did observe (this number includes companions),
several fields had data of poor quality owing to the same problems 
that we identified above.  These were TWA 2AB, 3AB, 5AB and 16AB.  
Thus of the 30 stars in 19 TWA systems listed by \citet{Webb99},
\citet{Sterzik99} (TWA $1-13$) and \citet{Zuckerman01a} (TWA $14-19$),
we analyzed the light curves for 16 stars in 12 systems\footnote{The
TWA has been expanded since our survey to include the star denoted 
TWA 20 by \citet{Reid03} for which membership of the TWA is disputed
by \citet{Song03}, and 5 stars (denoted TWA $21-25$) announced as
TWA members by \citet{Song03}.}.  

\begin{figure*}
\begin{center}
\includegraphics[scale=0.9]{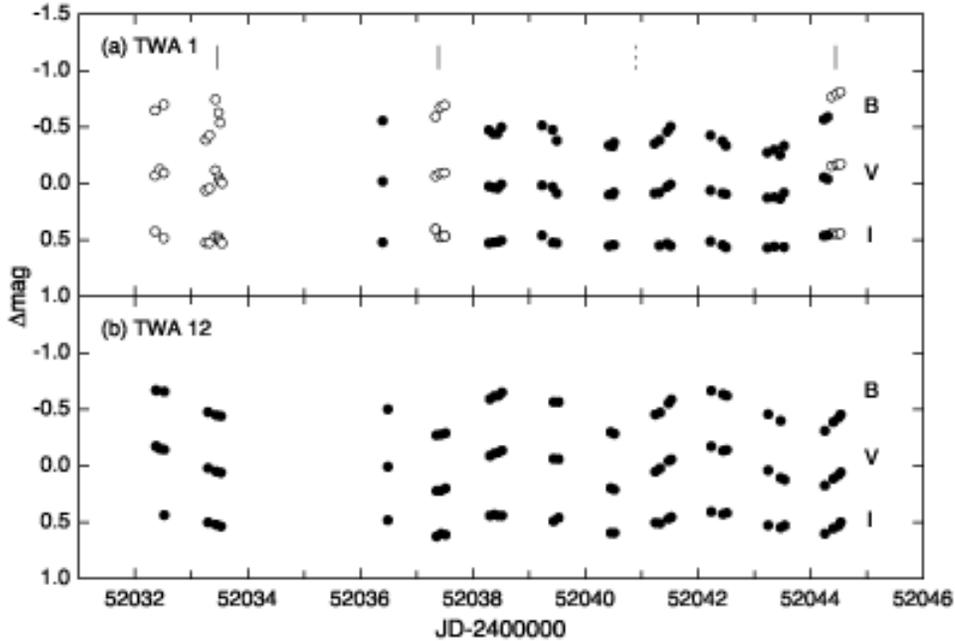}
\caption{{\it BVI\,} light curves for (a) TWA 1 = TW Hya and (b)
TWA 12; example light curves showing the temporal coverage of
our survey.  For TW Hya, short-lived unusually blue photometric 
excursions are indicated by the vertical bars that may be related 
to the veiling enhancements reported to occur on a $\sim 4$ d 
timescale (Alencar \& Batalha 2002).  The third bar near JD 
2452041 is shown as a dashed line since this event, if it occurred,
happened during the daytime.  Photometric data for TW Hya not
included in the light curve analysis are shown as open symbols; 
see Section 3.1 for discussion of the TW Hya light curve.}
\label{F1}
\end{center}
\end{figure*}

The production of the differential light curves for the TWA stars 
and their subsequent analysis using the Lomb-Scargle Fourier method 
follows that described by \citet{Lawson01} for their analysis of 
photometry of the late-type stars of the $\eta$ Cha cluster 
using the same telescope and instrumentation. For the TWA stars, 
with {\it VI\,} or {\it BVI\,} multi-colour light curves available 
for study, we concentrated our analysis on the $V$-band datasets 
which had superior signal-to-noise (S/N) ratio compared to the data 
obtained at the other photometric bands.  However, we still repeated
the analysis for the $B$- and $I$-band datasets, to confirm the 
period derived from the $V$-band datasets and to produce amplitudes 
for all of the observed colours.  We list in Table \ref{T1} the 
periods $P$ derived from the $V$-band datasets, the {\it BVI\,} 
amplitudes derived from the analysis of the individual colour curves 
(we list the peak-to-peak amplitude of the most-significant periodicity), 
and the S/N ratio of the $V$-band periodicity based upon measurement of
the noise level in the Fourier spectrum following pre-whitening of the 
original datasets.  In Fig. \ref{F1} we show two examples of our 
datasets: the {\it BVI\,} light curves for TW Hya and TWA 12.  
Both stars are clearly variable, with a periodicity 
of $\approx 3$ d duration being visible in the TWA 12 light curves.
These light curves also serve to demonstrate the temporal coverage 
achieved during our survey.  Most of the TWA stars were observed 
on as many epochs as TWA 12 ($\approx 30$), and only one star 
(TWA 6) was observed more often than the examples shown here.

We can address the reliability of our measured periods for the TWA
stars in several ways.  First, we appeal to our study of rotation 
periods in the $\eta$ Cha star cluster. \citet{Lawson01} observed 
these stars during 3 week observing runs in both 1999 and 2000, 
and found excellent agreement in the periods across the 2 observing
seasons.  Our measurements of the TWA stars were made at a 
higher frequency than the $\eta$ Cha stars which will reduce the 
risk of aliasing errors, thus we are confident about the periods 
listed for the TWA stars in Table \ref{T1}.  
Secondly, the period of 2.80 d measured for TW Hya (see Section 3.1
for more details of the analysis of the TW Hya light curve) agrees 
well with values of $P = 2.88$ d determined from {\it Hipparcos\,} 
photometry by \citet{Koen02} and $P = 2.85$ d determined from 
H$\beta$ veiling-corrected equivalent widths by \citet{Alencar02}.  
TW Hya is the only TWA star with a previously published period.  
We also note that TW Hya is a classical T Tauri (CTT) star with ongoing
mass accretion from its circumstellar disk \citep{Muzerolle00}.
Thus rotational variability in TW Hya is likely to be modulated
by accretion hotspots rather than cool starspot activity, and 
stochastic variations in the light curve owing to variable accretion
and flaring activity might translate into differences in the 
measured period. The other
TWA stars that we observed are non-accreting T Tauri stars and
their light curves will not be similarly affected.  Third, in Fig. 
\ref{F2} we compare our periods to another rotational measure; 
projected rotational velocities $v\sin i$ for the TWA stars taken 
mostly from the compilation of \citet{Reid03} except for TW Hya, 
for which we adopt $v\sin i = 5$ km\,s$^{-1}$ from \citet{Alencar02}.  
Reassuringly, Fig. \ref{F2} contains no unpleasant surprises, i.e.
there are no stars that have the unphysical combination of long
periods and high $v\sin i$ values.  Most of the stars are located 
within a narrow band between $P - v\sin i$ relations for a rotating
$M = 1$ M$_{\odot}$, $R = 1$ R$_{\odot}$ star seen at inclination 
angles $i = 10^{\circ}, 30^{\circ}$ and 90$^{\circ}$.  According to 
the PMS evolutionary tracks of Siess, Dufour \& Forestini (2000) these 
are roughly appropriate values for mass and radius for a late K-type
PMS star of age $\sim 10$ Myr.  For simplicity when calculating the 
relations, we assumed solid body rotation.  With the exception of 
TWA 8B, a low-mass star of spectral type M5, all of the TWA stars
plotted in Fig. \ref{F2} occupy a limited range of spectral types 
from K5 -- M2.5 \citep{Reid03}.  Thus the comparison shown in Fig.
\ref{F2} is reasonable.  As a further test, for TW Hya (indicated 
in Fig. \ref{F2} by its TWA number of `1') that has a spectral type 
of K7, we derive an inclination angle $i \approx 16^{\circ}$, in 
agreement with the value of $i = 18 \pm 10^{\circ}$ determined 
from an analysis of emission line profiles by \citet{Alencar02}.

\begin{table}
\centering
\caption{Rotation periods and amplitudes for TWA stars.}
\label{T1}
\begin{tabular}{@{}lccccc@{}}
\hline
     & Period & $\Delta B$ & $\Delta V$ & $\Delta I$ & S/N \\
Star &  (d)   &   (mag)    &   (mag)    &   (mag)    & ratio \\
\hline 
TWA 1   & 2.80 & 0.22 & 0.12 & 0.06 & 3 \\
TWA 6   & 0.54 & 0.54 & 0.49 & 0.27 & 30 \\
TWA 7   & 5.05 & ---  & 0.05 & 0.03 & 6 \\
TWA 8A  & 4.65 & ---  & 0.05 & 0.02 & 3 \\
TWA 8B  & 0.78 & ---  & 0.08 & 0.05 & 5 \\
TWA 9A  & 5.10 & ---  & 0.09 & 0.05 & 6 \\
TWA 9B  & 3.98 & ---  & 0.08 & 0.04 & 5 \\
TWA 10  & 8.33 & ---  & 0.14 & 0.08 & 7 \\
TWA 12  & 3.28 & 0.35 & 0.36 & 0.16 & 10 \\
TWA 13A & 5.56 & ---  & 0.21 & 0.13 & 5 \\
TWA 13B & 5.35 & ---  & 0.27 & 0.17 & 13 \\
TWA 14  & 0.63 & ---  & 0.11 & 0.05 & 25 \\
TWA 15A & 0.65 & ---  & 0.13 & 0.06 & 12 \\
TWA 15B & 0.72 & ---  & 0.05 & 0.02 & 8 \\
TWA 17  & 0.69 & ---  & 0.12 & 0.07 & 15 \\
TWA 18  & 1.11 & ---  & 0.07 & 0.05 & 13 \\
\hline
\end{tabular}
\end{table}

\section{Analysis of the light curves}

In describing in detail the photometric behaviour of the TWA stars,
we consider two groups of TWA members: the compliation of members
(TWA $1-13$) by \citet{Webb99} and \citet{Sterzik99}, and the several
stars (TWA $14-19$) added by \citet{Zuckerman01a}.  We have two 
important reasons for doing so.  First, these two groups appear to 
be spatially separate.  While the TWA stars listed by \citet{Webb99}
and \citet{Sterzik99} are distributed across an area $> 200$ deg$^{2}$,
most of the stars appear roughly proximate to TW Hya (galactic 
coordinates $\ell$, {\bf b} = $273^{\circ}$, $23^{\circ}$).  Three 
of the four members of the original association with {\it Hipparcos\,}
parallaxes, TW Hya, TWA 4A and TWA 9A, are located at very similar 
distances of $d = 56^{+8}_{-6}$ pc, $47^{+7}_{-6}$ pc and 
$50^{+7}_{-5}$ pc, respectively.  The fourth star, HR 4796 
(= TWA 11A), appears at $\ell$, {\bf b} = $300^{\circ}$, $23^{\circ}$
and lies at the greater distance of $d = 67^{+4}_{-3}$ pc.

The six TWA systems announced by \citet{Zuckerman01a} are centred
at $\ell$, {\bf b} $\sim$ 300$^{\circ}$, 15$^{\circ}$ and contain 
the star TWA 19A (= HD 102458) with an {\it Hipparcos\,} distance 
of $d = 104^{+18}_{-13}$ pc.  The relative proximity of these 
stars to HR 4796 (on the sky) lead \citet{Zuckerman01a} to consider
them `near HR 4796' but if they, as a group, have distances 
comparable to TWA 19A then they lie considerably `beyond' HR 4796.
They also have optical photometry that is fainter than most of the
TWA stars `near' TW Hya of similar spectra type, consistent with 
them having a distance $> 70$ pc.  (We elaborate and quantify this 
point in Section 4.2.)  The claimed three-dimensional extent of the 
TWA is thus immense; HR 4796 lies $\approx 20^{\circ}$ to the east
of TW Hya and is $\approx 10$ pc more-distant, while TWA 19A lies 
$\approx 20^{\circ}$ to the south of TW Hya and is $\approx 50$ pc
more-distant\footnote{Of the several TWA members announced by 
\citet{Song03}, TWA 22 is claimed to have a distance of only 22 pc 
based on colour-magnitude diagram placement.}. \citet{Song02} have 
even questioned the TWA membership of the TWA 19AB binary.  While 
they are clearly young stars owing to the detection of strong 
$\lambda$6707 Li I absorption and chromospheric activity in both 
TWA 19A and 19B \citep{Zuckerman01a}, the location of TWA 19A 
appears to place it within the LCC subgroup of the OSCA.  LCC 
subgroup stars with {\it Hipparcos\,} distances have an average 
distance of $\approx 120$ pc (range of 
$95-170$ pc); see table 4 of \citet{Mamajek02} for details.  
Fig. \ref{F3} shows TWA 19A (and the entire TWA $14-19$ group) 
superimposed on the confines of the LCC subgroup.  The placement 
of TWA 19A within the LCC subgroup would then distinguish the star
from the nearer TWA stars, and other nearby PMS groups such as the 
$\eta$ Cha and $\epsilon$ Cha clusters, which are thought of as 
dispersed OSCA groups \citep{Mamajek00, Feigelson03}.   (The nature
of the TWA 19AB system is further discussed in Section 4.2.)

\begin{figure}
\begin{center}
\includegraphics[scale=0.9]{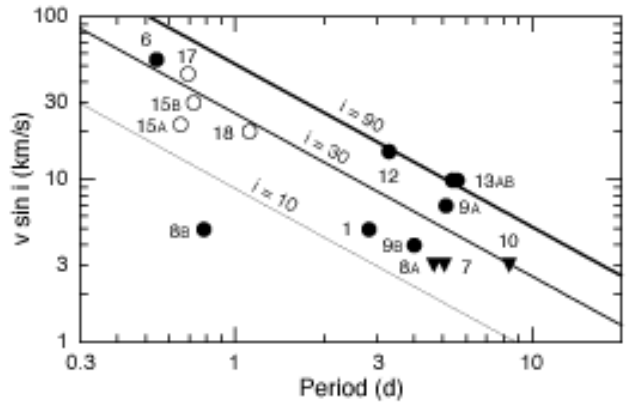}
\caption{
Rotation periods for TWA stars {\it versus\,} their projected
rotational velocity $v\sin i$.  TWA stars in the TWA $14-19$ group
are shown as open symbols.  Stars with upper-limit values for 
$v\,sin i < 3$ km\,s$^{-1}$ are shown as triangles.  The three 
relations are for a rotating $M = 1$ M$_{\odot}$, $R = 1$ R$_{\odot}$
solid body seen at inclination angles $i = 10^{\circ}, 30^{\circ}$
and 90$^{\circ}$.  Not all TWA stars with rotation periods have 
measured $v\sin i$ velocities, and {\it vice versa}.
}
\label{F2}
\end{center}
\end{figure}

\begin{figure}
\begin{center}
\includegraphics[scale=0.9]{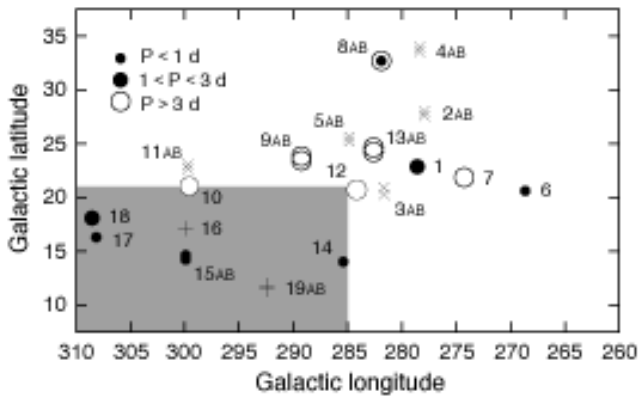}
\caption{
The galactic distribution of rotation periods for TWA members. 
The symbols representing rotation periods for the binaries TWA 
9AB, TWA 13AB and TWA 15AB are slightly offset for clarity.  
Stars denoted by crosses ($\times$) or pluses (+) are TWA stars 
without measured rotation periods.  The light grey region 
represents the northwestern extent of the LCC subgroup.
}
\label{F3}
\end{center}
\end{figure}

Secondly, a cursory evaluation of the periods listed in Table \ref{T1}
indicates that these two groups of TWA stars may be rotationally 
distinct.  For the TWA stars `near' TW Hya (TWA $1-13$) the group 
median is $P \approx 4.7$ d (range of $0.54 - 8.33$ d), whereas for 
the TWA stars lying `beyond' HR 4796 (TWA $14-19$) the group median 
is only $P \approx 0.7$ d (range of $0.63 - 1.11$ d).  As the range 
of spectral types for the stars in both groups with rotation periods
is similar, the TWA $14-19$ group will be expected to have higher 
rotational velocities than the TW Hya group.  This is strongly 
indicated in Fig. \ref{F2}.  Indeed, TWA 15A, 15B, 17 and 18 (shown 
as open symbols in Fig. \ref{F2}) straddle $v\sin i \approx 30$ 
km\,s$^{-1}$, whereas within the TWA $1-13$ group only TWA 6 has 
a higher $v\sin i$ velocity of 55 km\,s$^{-1}$. 

Both spatial and rotational distinctions are summaried graphically 
in Fig. \ref{F3}, where we plot the galactic distribution of TWA
$1-19$.  This figure not only shows the location of the TWA stars,
but the symbols for those stars with measured rotation periods are 
coded into three coarse period bins.  Most of the TWA stars near 
TW Hya are seen to have periods $> 3$ d, whereas two-thirds (4/6) 
of the stars with $P < 1$ d are associated with the TWA $14-19$ 
group of stars.  TWA stars without measured periods are denoted as 
crosses for stars in the TWA $1-13$ group, or as pluses for stars
in the TWA $14-19$ group.

Throughout the remainder of this paper, we proceed with the distinction
between the two groups of TWA stars: those that appear to be located 
`near' TW Hya (TWA $1-13$), and those that appear to be located `beyond'
HR 4796 (TWA $14-19$).  In the following subsections we discuss the 
light curves of the stars that we measured, and then in Section 4 
we discuss further the results of the period analysis.  We also adopt, 
other than for TW Hya (= TWA 1), HR 4796 (= TWA 11A) and HD 102458 
(= TWA 19A), the TWA number when discussing individual stars.  The 
{\tt SIMBAD\,} database provides a translation service from TWA 
number to other catalogue designations.  Common names for the 
brighter TWA stars are listed by \citet{Reid03}.

\subsection{TWA stars located `near' TW Hya}

\begin{figure*}
\begin{center}
\includegraphics[scale=0.85]{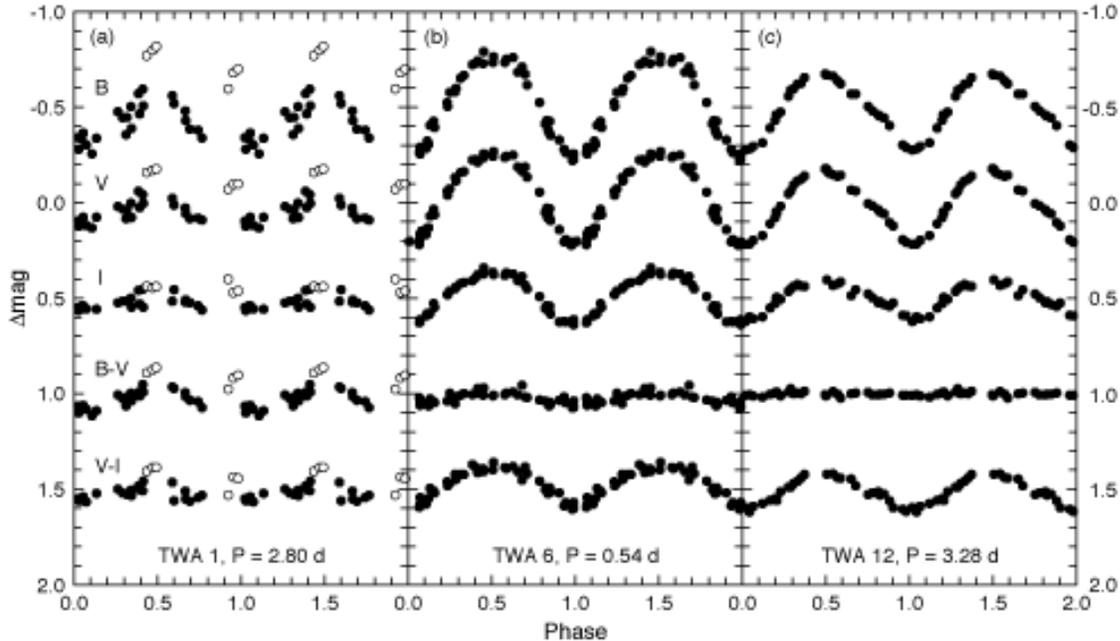}
\caption{
Phased {\it BVI\,} light and colour curves for (a) TWA 1 = TW Hya,
(b) TWA 6 and (c) TWA 12.  For TW Hya, only the photometry obtained 
between JD $2452036-45$ are phased, with the unusually blue 
photometric points that were removed from the periodic analysis 
shown as open symbols; see Section 3.1 for details.
}
\label{F4}
\end{center}
\end{figure*}

\begin{figure}
\begin{center}
\includegraphics[scale=0.86]{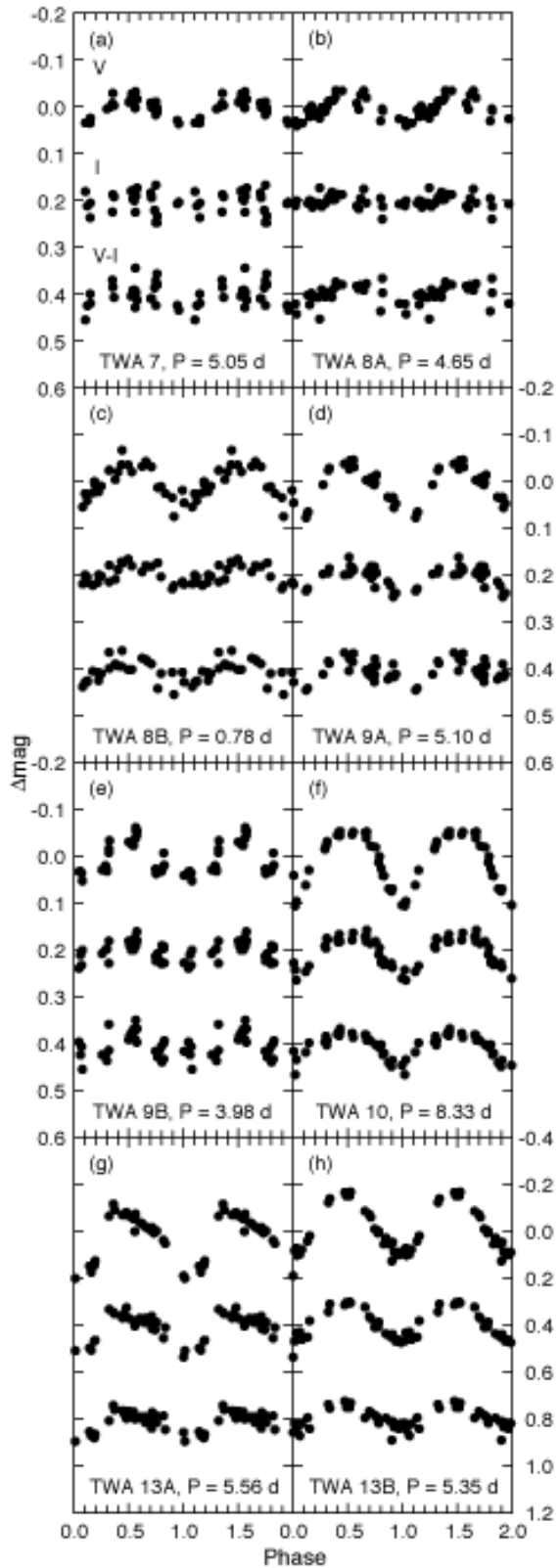}
\caption{Phased {\it VI\,} light and colour curves for (a) 
TWA 7, (b) TWA 8A, (c) TWA 8B, (d) TWA 9A, (e) TWA 9B, (f) 
TWA 10, (g) TWA 13A and (h) TWA13B.   Note that for TWA 13A 
and 13B the ordinate scale is decreased by a factor of 2 
compared to the other stars.}
\label{F5}
\end{center}
\end{figure}

Phased {\it BVI\,} light curves for TW Hya, TWA 6 and 12 are
shown in Fig. \ref{F4}, and phased {\it VI\,} light curves
for TWA 7, 8AB, 9AB, 10 and 13AB are shown in Fig. \ref{F5}.

For most of these T Tauri stars, the {\it VI\,} light curves 
appear to be reasonably regular over the 13-d observing window,
indicative of amplitude modulation due to cool starspots 
that are not evolving significantly during the course of the 
observations.  With the exception of TW Hya, discussed below,
these stars are WTT stars with low levels of optical emission 
and photospheric veiling, and no observed mass accretion 
observed from inner circumstellar disks that might
influence the photometry \citep{Webb99, Muzerolle00, Uchida04}.
The regularity of the WTT star light curves is particularly 
well-demonstrated in Fig. \ref{F4}, where the light curves 
for TWA 6 (Fig. \ref{F4}b) and TWA 12 (Fig. \ref{F4}c) are 
phase-wrapped over 22 and 4 cycles, respectively.  Several 
stars, e.g. TWA 7 (Fig. \ref{F5}a) show poor $I$-band light 
curves owing to these stars often being the brightest $I$-band 
source in the CCD field, and thus the local comparison stars 
in the CCD field used to define the differential $I$-band magnitudes 
were under-exposed compared to the TWA star.  A common feature 
of these light curves is that the ratio of the $V$- to $I$-band
amplitudes is $\approx 2$, therefore the $(V-I)$ colour amplitude 
is about half the $V$-band light amplitude; the measured amplitudes 
for these stars are listed in Table \ref{T1}.  A decrease in 
photometric amplitude with increasing wavelength is characteristic 
of the reduced contrast between the starspots and the surrounding 
photosphere; see \citet{Bouvier93}.  Most of these stars have
$V$-band amplitudes comparable to that observed in the 
similarly-aged $\eta$ Cha cluster \citep{Lawson01}, with
peak-to-peak amplitudes of $0.1-0.2$ magnitudes indicating
$\approx 10-20$ per cent differences in the $V$-band flux
between the most- and least-spotted hemispheres of the star.
The exceptions are the large amplitude variations seen in
TWA 6 and 12 where the peak-to-peak amplitudes indicate flux
differences at the $\approx 40-50$ per cent level during the
rotation cycle of these stars (an analogue in the $\eta$ Cha 
cluster is RECX 10; see Lawson et al. 2001).  The behaviour 
of the colour curves -- a negligible $(B-V)$ amplitude, but 
significant $(V-I)$ amplitude -- eliminates any prospect that 
the light curves are the result of an eclipsing binary system 
which would result in near-grey variations.  TWA 6 does not 
appear to be a binary according to \citet{Reid03}, while TWA 12 
remains a candidate binary star based upon one discrepant velocity 
(of only four measurements) obtained by \citet{Torres03}.

The exception within the TWA $1-13$ group of stars with measured
rotation periods is TW Hya itself.  The star is a CTT star with
ongoing moderate levels of mass accretion from its circumstellar
disk at the level of $\dot M \approx$ $5 \times 10^{-10}$ 
M$_{\odot}$\,yr$^{-1}$ \citep{Muzerolle00}.  The star shows 
a rich optical emission spectrum and strong blue and ultraviolet 
excess emission owing to the mass accretion.  All of these factors
might influence the light curve of the star.  Of all the TWA 
stars that we measured for rotation periods, the light curve 
for TW Hya proved to be the most challenging to interpret.  
The light curve (see Figs \ref{F1}a and \ref{F4}a) 
lacks the regularity seen in the WTT stars owing to rotational
modulation likely driven by accretion hotspots rather the passage
of cool slowly-evolving starspots.  In addition to rotational
variations, we appear to have resolved a second (pseudo-) 
periodic feature
in the light curve.  Compared to the general appearance of the
$B$-band light curve, with low-amplitude 2.80-d variations that
we ascribe to rotation (see below), we detected three short-lived
(timescale $\sim 1$ d) unusually blue excursions in the photometry
at JD 2452033.4, 2452037.3 and 2452044.4, respectively.  The 
interspacing of these features suggest a characteristic timescale 
of $\sim 3.7-4$ d.  If real, while the period is not strictly 
regular, another blue feature might have been expected near JD 
2452040.9 which was during the daytime in South Africa.  

We wonder if this feature is related to the $4.4 \pm 0.4$ d 
variations seen in the $B$-band veiling by \citet{Batalha02}, 
and in some H$\alpha$ and H$\beta$ line intensity and equivalent 
width datasets by \citet{Alencar02}.  Although less visible in 
$V$- and $I$-band, these features have a major impact on the 
Fourier analysis of the light curves, and prevented identification 
of any statistically-significant periodicity.  Reasoning that 
these features might not be related to stellar rotation, we 
removed the data points around the times of the blue features 
from the periodic analysis.  We also considered only the continuous 
series of observations between JD $2452036-45$, as few epochs
then remained in the JD $2452032-33$ interval.  The remaining 
data points are shown as filled symbols in Fig. \ref{F1}(a).  
Fourier analysis of these observations recovered the $P = 2.80$ d
periodicity that we list in Table \ref{T1}.  This period is visible 
in all the light and colour curves; see Fig. \ref{F4}(a) where we 
phase-plot the data obtained between JD $2452036-45$.  The low S/N
ratio of 3 for the detection of the 2.80-d period in the $V$-band 
is evidence for the variations being caused by hotspots, resulting 
in irregular amplitudes.  As we discussed in Section 2, periodicities 
of comparable value have been recovered in other datasets, in 
particular a 2.88-d period found in the {\it Hipparcos\,} light 
curve for TW Hya \citep{Koen02}.   The ratio of the $3.7-4.8$ d 
period to the $2.8-2.9$ d period is $1.5 \pm 0.2$.  We speculate 
that a 3:2 resonance exists between the $\approx 2.8$ d rotational 
period of the star and the in-fall of accreted material on a 
$\sim 4$ d cycle, causing an enhancement of photospheric veiling
which drives the $B$-band photometry, and a strengthening of 
optical line emission.

The photometric amplitude of TW Hya is near the average of that 
seen in the other TWA stars we measured.  This is in contrast to
observations of other CTT stars that show high levels of optical 
variability; see, e.g. \citet{Bouvier93}.  The relatively low 
level of variability seen is likely a consequence of the near 
pole-on alignment of the rotation axis of TW Hya, for which 
\citet{Alencar02} derived an inclination $i = 18 \pm 10^{\circ}$.

\subsection{TWA stars located `beyond' HR 4796}

\begin{figure*}
\begin{center}
\includegraphics[scale=0.86]{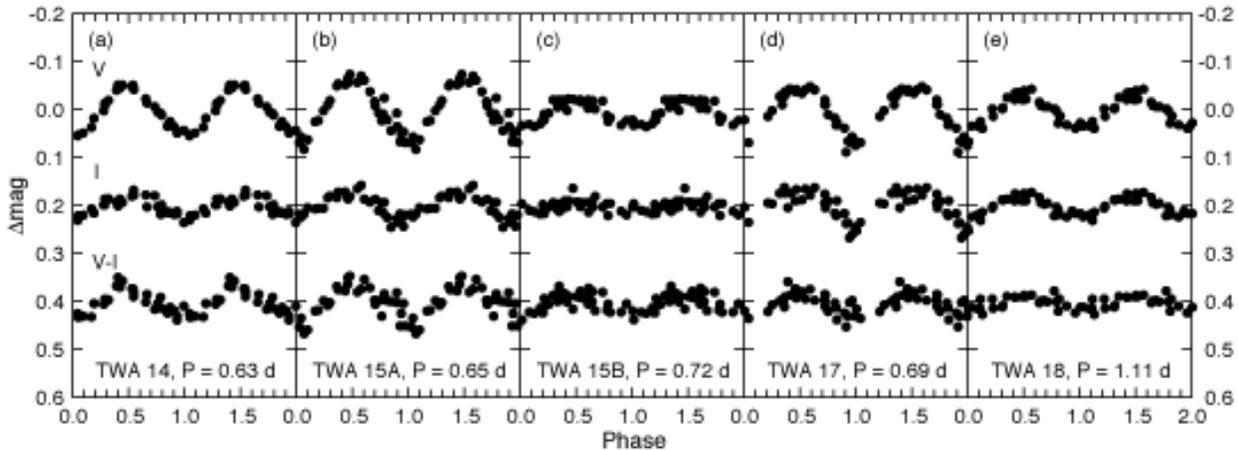}
\caption{Phased {\it VI\,} light and colour curves for (a) 
TWA 14, (b) TWA 15A, (c) TWA 15B, (d) TWA 17 and (e) TWA 18.}
\label{F6}
\end{center}
\end{figure*}

Phased {\it VI\,} light curves for TWA 14, 15AB, 17 and 18 
are shown in Fig. \ref{F6}.  Owing to the short periods of 
$0.63-1.11$ d detected in these stars and the number of 
observations made, the phased light curves appear well-filled 
with the photometry wrapped over $11-19$ rotation cycles.  
All five stars showed regular light curves over the course 
of the observations, as is expected for WTT stars with light 
curves rotationally modulated by cool starspots.

\section{Discussion}

\subsection{Statistical comparisons}

\begin{figure}
\begin{center}
\includegraphics[scale=0.85]{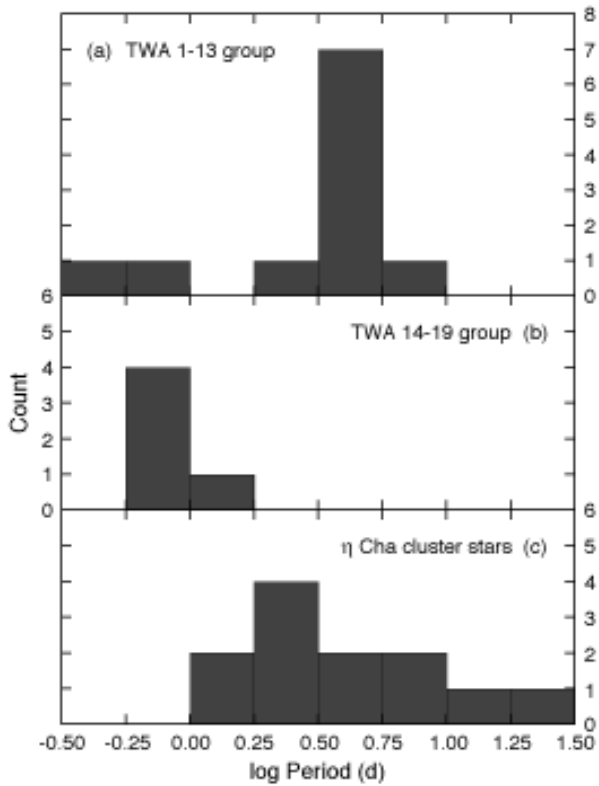}
\caption{Rotation period distributions for (a) TWA $1-13$
group stars located `near' TW Hya, (b) TWA $14-19$ group 
stars located `beyond' HR 4796, and (c) $\eta$ Cha cluster 
members.}
\label{F7}
\end{center}
\end{figure}

To assess the significance of the short rotation periods found 
for the TWA $14-19$ group, we compared these periods to 
those obtained for the TWA $1-13$ group associated with 
TW Hya, and also with the periods measured for stars in the $\eta$
Cha cluster by \citet{Lawson01, Lawson02}.  The distribution of 
rotation periods for these three groups of PMS stars is shown in
Fig. \ref{F7}.  We performed a Mann-Whitney non-parametric
test to statistically evaluate the differences betwen the three 
samples.  This test makes no underlying assumptions concerning 
the distribution of the datasets, e.g. there is no assumption 
that samples are normally distributed with the same variances.  
In comparing pairs of samples, we found the probability ${\cal P}$
that the two groups have been drawn from the same parent population
are for: (i) the TW Hya group compared to the $\eta$ Cha group;
${\cal P} = 0.93$, (ii) the TW Hya group compared to the TWA 
$14-19$ group; ${\cal P} = 0.013$, and (iii) the $\eta$ Cha 
group compared to the TWA $14-19$ group; ${\cal P} = 0.0003$.  
Thus at the $\approx 2 \sigma$ level, the TW Hya group and the 
$\eta$ Cha cluster stars show a similar distribution of rotation 
periods.  This result is not surprising as both groups of stars 
are believed to be similarly aged at $t = 8-10$ Myr, and they 
are both outlying populations of the OSCA \citep{Mamajek00}.  
However, the comparison with the TWA $14-19$ group of stars 
presents a very different result, with $\approx 3 \sigma$ 
(or greater) likelyhood that the periods of these stars 
are {\it not\,} drawn from the same parent population as that 
for the TW Hya or $\eta$ Cha groups.

\subsection{The distance to the TWA 14 -- 19 group}

\citet{Zuckerman01a} noted that the TWA $14-19$ group of stars 
were `somewhat fainter at optical and infrared wavelengths than
previously known members of comparable spectral type'.  Since 
most of the stars that we measured for rotation periods occupy a
limited range of spectral types of K5 -- M2.5, with the single
exception being the M5 star TWA 8B, we quantified the difference 
in brightness by considering the average 2MASS $J$-band flux for
the stars in the two groups for which we measured rotation periods.
(By only comparing the stars with rotation periods, we ensured
we were dealing with the photometry of individual stars, not the
merged photometry of binaries.)  We chose the 2MASS survey as it 
provided a uniform source of photometry and $J$-band fluxes are
a reasonable surrogate for stellar luminosity in this spectral
range.  For dwarf stars of spectral type K5 -- M2.5, the bolometric 
correction required to correct $J$-band magnitudes into stellar
luminosities varies by only 0.15 magnitudes; see fig. 5 of 
\citet{Lawson96}.

For the 10 stars in the TWA $1-13$ group with rotation periods 
listed in Table \ref{T1}, excluding the M5 star TWA 8B, we find 
$\overline{J} = 8.55 \pm 0.06$ (standard error; se).  For the 5
stars in the TWA $14-19$ group listed in Table \ref{T1}, we find 
$\overline{J} = 9.92 \pm 0.10$ (se).  If both groups are coeval
and unreddened, then a $\approx 1.4$ magnitude difference 
corresponds to the TWA $14-19$ group being located $1.9 \times$
more-distant than the TWA $1-13$ group.  This places TWA $14-19$
at an average distance of $\approx 105$ pc, in comparison with the 
average of {\it Hipparcos\,} distances to TW Hya, TWA 4A, TWA 9A
and HR 4796 of $\overline{d}$ = 55 pc.  A similar result can be
gleaned from table 2 of \citet{Song03}, who derived distances to
the TWA stars from colour-magnitude diagram placement, assuming 
that the stars were coeval.  Of course, if the two groups are not
coeval then the above arguments need to be modified; a younger
TWA $14-19$ group would be more-distant owing to the higher 
luminosity of younger PMS stars, and accordingly an older 
less-luminous TWA $14-19$ group would be located closer. However,
the two groups cannot be codistant unless the TWA $14-19$ stars, 
excluding HD 102458 with an astrometric distance, have an age of
$\sim 100$ Myr. This can be ruled out owing to the detection of 
largely-undepleted lithium in these stars \citep{Zuckerman01a}.

An average distance of $\approx 105$ pc to the TWA $14-19$ group
is consistent with the {\it Hipparcos\,} distance to HD 102458
of $d = 104^{+18}_{-13}$ pc.  However, H-R diagram placement 
of HD 102458 and its companion TWA 19B, when compared to the 
PMS tracks of \citet{Siess00}, suggests that the two stars are 
not coeval\footnote{To locate HD 102458 in the H-R diagram, 
we converted {\it Hipparcos}/Tycho $B_{T}$, $V_{T}$ photometry 
to the Johnson/Cousins {\it BV\,} system using transformation 
equations given by \citet{Mamajek02}.  We adopted a spectral 
type of G5 with no reddening from \citet{Zuckerman01a}, or F9
with $A_{V} = 0.3$ magnitudes from \citet{Mamajek02}. For TWA 19B 
we adopted {\it VI\,} photometry from \citet{Reid03}, and a
spectral type of K7 or M0 with no reddening.  We then used the 
dwarf bolometric correction and temperature sequences given by
\citet{Kenyon95}, which are appropriate for older PMS populations 
\citep{Mamajek02, Lyo04}.}.  For HD 102458, our derived age 
is $\approx 17$ Myr, in agreement with values of $15-19$ Myr 
derived using various PMS grids by \citet{Mamajek02}.  TWA 19B
appears to have an age of only $\approx 5$ Myr. The apparent age
discrepancy can only be explained if the stars are unrelated or
if TWA 19B has elevated luminosity, as the difference is too 
large to be explained by uncertainties in the PMS tracks and the
reddening towards these stars is low.  From proximity and 
kinematic arguments the two stars are clearly related; TWA 19B 
resides $\approx 37$ arcsec east of HD 102458 and they are 
comoving; HD 102458 has an {\it Hipparcos\,} proper motion of 
$\mu_{\alpha}$, $\mu_{\delta}$ = $-33.7 \pm 1.1$, 
$-9.1 \pm 1.1$ mas\,yr$^{-1}$ (se), whereas TWA 19B has a UCAC2 
proper motion of $\mu_{\alpha}$, $\mu_{\delta}$ = $-35.6 \pm 4.8$,
$-7.5 \pm 4.6$ mas\,yr$^{-1}$ (se).  The apparent youthful age 
of TWA 19B can be reconciled with that of HD 102458 if it is 
an undocumented binary.  Assuming that TWA 19B is a binary with
equal luminosity components, they then have an age of $\approx 15$
Myr.  Merging the above age estimates for HD 102458/TWA 19B, we 
adopt $t = 17 \pm 2$ Myr as the age of the system.

An age of $t \approx 17$ Myr for the HD 102458 system is encompassed
within the distribution of ages derived from H-R diagram placement 
for LCC subgroup PMS stars and PMS candidates of $22.5 \pm 7.8$ 
Myr (1 $\sigma$) by \citet{Mamajek02}, using the PMS grids of 
\citet{Siess00}.  Both HD 102458 and TWA 19B have proper motions
consistent with LCC membership; see table 1 of \citet{Mamajek02}.
The spatial location, age and proper motions of HD 102458/TWA 19B
together present a convincing case for a LCC subgroup origin for
the HD 102458 binary. 

Other TWA $14-19$ group members with UCAC2 proper motions similarly
present a strong case for LCC subgroup membership\footnote{TWA 
16 has UCAC2 proper motions of $\mu_{\alpha}$, $\mu_{\delta}$ = 
$-53.3 \pm 5.2$, $-19.0 \pm 5.2$ mas\,yr$^{-1}$ (se), discordant
with LCC subgroup membership \citep{Mamajek02}.  However, TWA 16
is a binary with a separation of $\approx 0.67$ arcsec and a brightness
ratio $\sim 0.9$ \citep{Zuckerman01a}.  The UCAC2 catalogue does
not resolve the components, and the proper motion is likely in 
error.}.  TWA 14 is located near HD 102458 (see Fig. \ref{F3})
and has similar proper motions of $\mu_{\alpha}$, $\mu_{\delta}$ = 
$-43.4 \pm 2.6$, $-7.0 \pm 2.4$ mas\,yr$^{-1}$ (se).  TWA 18 has 
proper motions of $\mu_{\alpha}$, $\mu_{\delta}$ = $-29.0 \pm 5.2$,
$-21.2 \pm 5.2$ mas\,yr$^{-1}$ (se), comparable to LCC stars at
a similar Right Ascension \citep{Mamajek02}.

If the age of the HD 102458 system reflects that of other stars 
in the TWA $14-19$ group, then these stars are clearly not coeval 
with the TWA $1-13$ group which has a derived age of only $8-10$
Myr from H-R diagram placement; see e.g. fig. 3 of \citet{Webb99}.  
The average photometric distance calculated above for the TWA $14-19$
group would then reduce to $d \approx 90$ pc, which locates the group 
$\sim 35$ pc beyond stars in the TWA $1-13$ group while still 
remaining consistent with the astrometric distance to HD 102458
and other members of the LCC subgroup.

\section{Summary}

Our photometric study of 16 members of 12 TWA systems found 
that stars measured in the TWA $1-13$ group led by TW Hya have 
a distribution of rotation periods (median period $P = 4.7$ d)
that differs significantly from that measured for stars in the 
TWA $14-19$ group (median period $P = 0.7$ d).  The two rotation 
period distributions cannot be reconciled at the 3-$\sigma$ 
significance level; see Sections 3 and 4.1 for details of the 
light curve analysis.  Summarizing our discussions presented in
Section 4.2, the TWA $14-19$ group likely resides at an average 
distance of $\approx 90$ pc, coincident with the {\it Hipparcos\,} 
distance to HD 102458 (= TWA 19A) of $d \approx 104$ pc 
and the location of the near boundary of the LCC subgroup at 
$d \approx 95$ pc.  HD 102458, TWA 14, 18 and 19B also have proper
motions consistent with the LCC subgroup.  From the derived age
of HD 102458, these stars might also be $\approx 8$ Myr older 
than the TW Hya group ($\approx 17$ Myr {\it versus\,} $8-10$ Myr),
compatible with the ages derived for LCC subgroup PMS stars 
and candidate members by \citet{Mamajek02}.

The difference in the rotation period distributions of the two
groups of TWA stars may be further evidence that the TWA $14-19$
group have a greater age than the TWA $1-13$ group; these stars 
may have had an additional $\approx 8$ Myr to undergo rotational 
spin-up following dissipation of their circumstellar discs.  
The disc dispersal process in low mass PMS stars might even be 
accelerated within the environs of an OB-star association
compared to that for a dispersed PMS group such as TWA $1-13$, 
e.g. winds from resident O- and B-type stars or an aggressive 
dynamical environment within the association might prematurely 
disrupt their discs.

We conclude that the two groups of TWA stars, TWA $1-13$ and TWA
$14-19$, are spatially- and rotationally-distinct populations of 
PMS stars that might also differ in age by a factor of $\approx 2$.
The availability of rotation periods has been a useful tool to help
distinguish these two populations.  We therefore do not consider 
that the stars denoted TWA $14-19$ represent an extension of the 
original TWA as proposed by \citet{Zuckerman01a}.  Most likely 
they are representive of the population of low mass PMS stars 
still physically associated with the LCC subgroup of the OSCA.

\section*{Acknowledgments}  

We thank the SAAO Time Allocation Committee for the generous award
of telescope time towards this and other projects related to the 
study of the nearest PMS populations. We also thank Rich Webb for 
providing us with pre-publication identifications for TWA $14-19$,
and Eric Feigelson for useful discussions concerning this work.  
We made extensive use of the {\tt SIMBAD} and {\tt VizieR} databases 
held at the Centre de Donn\'ees astronomiques de Strasbourg.  WAL's 
research is supported by UNSW@ADFA Faculty Research Grants and 
Special Research Grants.  LAC acknowledges support from the
National Research Foundation and the University of Cape Town.

\bsp
\label{lastpage}
\end{document}